\documentclass[10pt,a4paper]{article}
\usepackage[utf8]{inputenc}
\usepackage[english]{babel}
\usepackage{url}
\usepackage{hyperref}
\usepackage{amsmath}
\usepackage{amsfonts}
\usepackage{amssymb}
\usepackage{graphicx}
\usepackage{float}
\usepackage{lipsum}
\usepackage{multicol}
\usepackage{epstopdf}
\usepackage{xcolor}
\usepackage{algorithm}
\usepackage{cite}
\usepackage{amsmath,amssymb,amsfonts}
\usepackage{algpseudocode}
\usepackage{textcomp}
\usepackage{xcolor}
\usepackage{blindtext}
\usepackage{subfig}
\usepackage{cuted}
\usepackage{physics}
\usepackage{xcolor}
\usepackage[toc,page]{appendix}
\usepackage[long]{optidef}
\usepackage[detect-none]{siunitx}
\usepackage{multirow}
\usepackage{array}

\definecolor{azul}{rgb}{0.0, 0.53, 0.74}
\usepackage[left=2.00cm, right=2.00cm, top=2.00cm, bottom=2.00cm]{geometry}
\author{Aprendiendo \LaTeX\,}
\title{Mi primer paper}

\begin{document}
	\vspace{7.5mm}
	
	\begin{center}
		{\Large \textbf{Notes on Degeneracy and Robustness}}\\
		\vspace{2mm}
		{\large Indrakshi Dey$^{1}$, and Nicola Marchetti$^{2}$}\\
		\vspace{7.5mm}
		$^1$\textit{Walton Institute, South East Technological University, Waterford, Ireland} \\
		$^2$\textit{Trinity College Dublin, Dublin, Ireland}
	\end{center}







\section*{Degeneracy}\label{sec1}

Degeneracy is the ability of structurally different elements to perform the same function or yield the same output under certain constraints. In contrast to redundancy, which implies identical backups, degeneracy allows  diverse components to step in and perform the same or similar role. Let \( \mathcal{E} = \{ e_1, e_2, \dots, e_n \} \) is a set of system elements (nodes, subsystems, base-station, etc.), \( \mathcal{F} = \{ f_1, f_2, \dots, f_m \} \) is a set of functions/tasks to be performed , and \( \phi: \mathcal{E} \times \mathcal{F} \rightarrow \{0, 1\} \) is a binary mapping function:
  \[
  \phi(e_i, f_j) = \begin{cases}
  1 & \text{if } e_i \text{ can perform } f_j \text{ with acceptable quality} \\
  0 & \text{otherwise}
  \end{cases}
  \]
Now, we can define a \emph{Structural Dissimilarity Function}
\begin{align}
    D(e_i, e_k) > \delta \quad \text{implies } e_i \text{ and } e_k \text{ are structurally distinct}
\end{align}
where $\delta$ can be derived from a) Graph distance, b) Hamming distance in encoding, c) Architectural/algorithmic difference or d) Distance in feature space For a function \( f_j \), we define degeneracy as the number of structurally different elements that can perform it:
\begin{align}
    \mathcal{D}(f_j) &= | \{ (e_i, e_k) \mid \phi(e_i, f_j) \nonumber\\
    &= \phi(e_k, f_j) = 1 \land D(e_i, e_k) > \delta \} |
\end{align}
where $\mathcal{D}(f_j)$ is coined as \emph{Degeneracy Score}. A high \( \mathcal{D}(f_j) \) refers to many non-identical elements that can achieve the same outcome. If \( \mathcal{D}(f_j) = 0 \), such a condition refers to no degeneracy and only identical components perform the function. Degeneracy is not redundancy, but it is flexibility through diversity. Mathematically, it is about mapping multiple distinct elements into the same function. In a degenerate system, failure in one part can be compensated by others not structurally linked. System functions are distributed within the system itself or the entire network. This renders faster and more adaptive recovery. Below we define and formulate several novel metrics for resource fungibility to address robustness in networks (static/mobile/dynamic) \cite{1,2}.

\section*{Degeneracy-Weighted Path Robustness (DWPR)}\label{sec2}

The goal is to evaluate the ability of functionally equivalent but structurally diverse paths to sustain communications. Let \( P_{sd} = \{P_1, P_2, \dots, P_n\} \) denotes all valid paths between nodes \( s \) and \( d \), \( M(P_i) \) is the set of modes/resources used by path \( P_i \), \( \mathcal{M} = \{M(P_i)\} \) is the set of unique* mode/resource combinations, and \( Q(P_i) \) is a QoS function, like, throughput, delay, signal-to-interference-plus-noise ratio (SINR). Then we can define;
\begin{align}
    \text{DWPR}(s, d) = \frac{1}{|\mathcal{M}|} \sum_{P_i \in P_{sd}} \textbf{1}[Q(P_i) \geq \theta] \cdot D(M(P_i))
\end{align}
where \( D(M(P_i)) \) is the structural dissimilarity of that path to others (can be a graph distance or mode divergence metric) and \( \theta \) is the QoS threshold. \emph{Note} - Higher DWPR refers to more diverse paths provide acceptable performance which corresponds to stronger path-level robustness.

Now, let us consider a multi-modal communication network modeled as a labeled undirected graph, $G = (V, E, \mu)$, where \( V \) is the set of nodes (devices, relays, satellites, etc.), \( E \subseteq V \times V \) is the set of edges representing communication links and \( \mu: E \rightarrow \mathcal{M} \times \mathbb{R}_+ \times \mathbb{R}_+ \) is an edge labeling function given by, $\mu(e) = (\text{mode}, \text{latency}, \text{bandwidth})$. Let \( s, d \in V \) be the source and destination pair, \( \mathcal{P}_{sd} \) is a set of all simple paths from \( s \) to \( d \). If a path \( P = (v_1, v_2, \dots, v_k) \in \mathcal{P}_{sd} \) exists, we define QoS-valid paths as:
\begin{align}
    \mathcal{P}^{\text{valid}}_{sd} = \left\{ P \in \mathcal{P}_{sd} \;\middle|\; \sum_{(u,v)\in P} \lambda_{uv} \leq \Lambda, \quad \min_{(u,v)\in P} \beta_{uv} \geq \beta \right\}
\end{align}
where \( \Lambda \) and \( \beta \) are application-specific thresholds. 

Let us define \( \mathcal{M}(P) \) as the set of communication modes used in path \( P \in \mathcal{P}^{\text{valid}}_{sd} \) and the associated dissimilarity function as \( D(P_i, P_j) \), defined as,
\begin{align}
    D(P_i, P_j) = 1 - \frac{|\mathcal{M}(P_i) \cap \mathcal{M}(P_j)|}{|\mathcal{M}(P_i) \cup \mathcal{M}(P_j)|} \quad \text{(Jaccard dissimilarity)}
\end{align}
Here Jaccard dissimilarity is defined as how different two sets are. For two sets $A$ and $B$, the Jaccard similarity is given by $J(A,B) = |(A \cap B)(A \cup B)|$ and dissimilarity as,
\begin{align}
    D_J(A,B) = 1 - J(A,B) = \frac{|A \cup B| - |A \cap B|}{|A \cup B|}
\end{align}
If $D_J(A,B) = 0$, the sets are identical, $D_J(A,B) = 1$, the sets have no elements in common and $D_J(A,B) \in (0,1)$ is the degree of dissimilarity.

Now let us integrate path quality, reliability and entropy of mode usage across paths. Let \( \mathcal{P}^{\text{valid}}_{sd} \) be the set of QoS-valid paths between source \( s \) and destination \( d \). For each path \( P_i \in \mathcal{P}^{\text{valid}}_{sd} \), let us define \( \mathcal{M}(P_i) \) as the set of modes used and \( Q(P_i) \in \mathbb{R}_+ \) as a composite quality score, like,
\begin{align}
     Q(P_i) = \frac{\min(\beta)}{\sum \lambda + \text{hop count}}
\end{align}
Let us define a probability distribution over paths based on quality,
\begin{align}
    \mathbb{P}(P_i) = \frac{Q(P_i)}{\sum_{P_j \in \mathcal{P}^{\text{valid}}_{sd}} Q(P_j)}
\end{align}
where \( \mu(P_i) \) is the vector of frequency of each mode (normalized). Lets use mode entropy and Jensen-Shannon divergence (JSD) to get robustness from mode diversity:
\begin{align}
    \text{DWPR}^*(s, d) = H_{\text{mode}} + \sum_{i,j} \mathbb{P}(P_i)\mathbb{P}(P_j) \cdot \mathcal{J}(\mu(P_i), \mu(P_j))
\end{align}
where \( H_{\text{mode}} \) is the entropy of mode distribution over all valid paths, $\mathcal{J}$ is the Jensen-Shannon Divergence (JSD) \cite{3}  and it is a method used to measure the similarity between two probability distributions. It is based on Kullback-Leiber divergence (KL divergence) but is symmetric and always finite, make it more stable and practical for many applications.

Given two probability distributions \( P \) and \( Q \), the Jensen-Shannon divergence is defined as:
\begin{align}
    \text{JSD}(P \| Q) = \frac{1}{2} \text{KL}(P \| M) + \frac{1}{2} \text{KL}(Q \| M)
\end{align}
where $M = \frac{1}{2}(P + Q)$ and \( \text{KL}(P \| M) \) is the Kullback-Leibler divergence of \( P \) from \( M \) given by, $\text{KL}(P \| M) = \sum_x P(x) \log \frac{P(x)}{M(x)}$.

\section*{Functional Substitution Score (FSS)}

The goal is to measure how many distinct subsystems or algorithms can take over a failed components tasks. Let \( F \) be a function (e.g., data aggregation, routing, compression), \( \mathcal{E}_F = \{e_i\} \) is the set of elements that can perform \( F \) and \( D(e_i, e_j) \) is the structural dissimilarity between components. Then we can formulate,
\begin{align}
    \text{FSS}(F) = \frac{1}{|\mathcal{E}_F|(|\mathcal{E}_F| - 1)} \sum_{i \neq j} \mathbf{1}[D(e_i, e_j) > \delta]
\end{align}
This measures the functional degeneracy richness. Lets now apply it to a network. Let \( \mathcal{E}_F = \{e_1, e_2, \dots, e_n\} \) are the elements that can perform function \( F \) and \( D(e_i, e_j) \) is the structural dissimilarity between elements, like hardware, software, mode. If
\[
\mathbf{1}_{\delta}(e_i, e_j) = 
\begin{cases}
1 & \text{if } D(e_i, e_j) > \delta \\
0 & \text{otherwise}
\end{cases}
\]
then we can write,
\begin{align}
    \text{FSS}(F) = \frac{1}{n(n - 1)} \sum_{\substack{i, j \\ i \neq j}} \mathbf{1}_{\delta}(e_i, e_j)
\end{align}
If $\text{FSS} = 1$, all substitutes are structurally diverse and the resultant system is a resilient system.

Now let us consider each \( e_i \) has a structural embedding vector \( \vec{s}_i \in \mathbb{R}^d \), a functional capacity expressed as \( C(e_i) \) and real-time load given by \( L(e_i) \). Let us define pairwise structural divergence $D(e_i, e_j) = \|\vec{s}_i - \vec{s}_j\|_2$ and a weighted substitution matrix as 
\begin{align}
    W_{ij} = \frac{\min(C(e_i), C(e_j))}{1 + |L(e_i) - L(e_j)|} \cdot D(e_i, e_j)
\end{align}
In this case, we can arrive at the enhanced Functional Substitutional Score, $\text{FSS}^*(F)$
\begin{align}
    \text{FSS}^*(F) = \frac{1}{n(n-1)} \sum_{\substack{i \neq j}} \mathbf{1}_{D > \delta}(e_i, e_j) \cdot W_{ij}
\end{align}
Weights are structurally diverse substitutes more when they are functionally aligned and load-balanced.

\section*{Algorithmic Resilience Quotient (ARQ)}

The goal is to measure how many structurally different algorithms produce functionally equal outputs. Let \( \mathcal{A} = \{A_1, \dots, A_n\} \)be the set of candidate algorithms that can solve task \( T \), \( \vec{P}(A_i) \) is the performance vector of algorithm \( A_i \), \( D(A_i, A_j) \) is the structural dissimilarity between algorithms, $\epsilon$ is the functionality similarity threshold and $\delta$ is the structural diversity threshold. Let us also define,
\[
\mathbf{1}_{\epsilon, \delta}(i, j) = 
\begin{cases}
1 & \text{if } \|\vec{P}(A_i) - \vec{P}(A_j)\| \leq \epsilon \land D(A_i, A_j) > \delta \\
0 & \text{otherwise}
\end{cases}
\]
then we can define ARQ as
\begin{align}
    \text{ARQ}(T) = \frac{1}{n(n - 1)} \sum_{\substack{i, j \\ i \neq j}} \mathbf{1}_{\epsilon, \delta}(i, j)
\end{align}
Now let us use it for networks and use functional similarity kernels, performance envelope and algorithmic tree distance. Now, let \( A_i \) is an algorithm with performance vector \( \vec{P}(A_i) \) and \( \vec{S}(A_i) \) as structural representation (e.g., parse tree, code embedding).

Gaussian kernel similarity (also known as the Radial Basis Function (RBF) kernel) is a popular way to measure similarity between two vectors in a continuous space. Given two vectors \( \mathbf{x} \) and \( \mathbf{y} \), the Gaussian kernel is defined as, $K(\mathbf{x}, \mathbf{y}) = \exp\left(-\frac{\|\mathbf{x} - \mathbf{y}\|^2}{2\sigma^2}\right)$ or equivalently $K(\mathbf{x}, \mathbf{y}) = \exp\left(-\gamma \|\mathbf{x} - \mathbf{y}\|^2\right)$ where \( \|\mathbf{x} - \mathbf{y}\|^2 \) is the squared Euclidean distance between \( \mathbf{x} \) and \( \mathbf{y} \), \( \sigma \) (or \( \gamma = \frac{1}{2\sigma^2} \)) controls the width of the Gaussian — how quickly similarity drops off. When \( \mathbf{x} = \mathbf{y} \), \( K(\mathbf{x}, \mathbf{y}) = 1 \) (maximum similarity). As \( \mathbf{x} \) and \( \mathbf{y} \) move further apart, the similarity decays exponentially. The kernel is always in the range \( (0, 1] \).

Let us use Gaussian kernel similarity in performance space,
\begin{align}
    K_P(i, j) = \exp\left(-\frac{\|\vec{P}(A_i) - \vec{P}(A_j)\|^2}{2\sigma^2}\right)
\end{align}
and use tree-edit distance or cosine dissimilarity in structure:
\begin{align}
    D_S(i, j) = 1 - \frac{\langle \vec{S}(A_i), \vec{S}(A_j) \rangle}{\|\vec{S}(A_i)\| \|\vec{S}(A_j)\|}
\end{align}
In this case, we can then formulate enhanced ARQ as,
\begin{align}
    \text{ARQ}^*(T) = \frac{1}{n(n-1)} \sum_{\substack{i \neq j}} K_P(i, j) \cdot D_S(i, j)
\end{align}
which is the combination of functional similarity and structural difference continuously. 

It is noteworthy that the tree-edit distance can be mathematically defined om many ways  in the context of structural divergence between graphs represented by feature vectors like degree histograms, graphlet counts, etc. Let \( \mathbf{x} = (x_1, x_2, \ldots, x_n) \) and \( \mathbf{y} = (y_1, y_2, \ldots, y_n) \) be two feature vectors. The distance metric can be defined using any of the followings;
\begin{itemize}
    \item Euclidean Distance - $d_{\text{euclidean}}(\mathbf{x}, \mathbf{y}) = \sqrt{ \sum_{i=1}^n (x_i - y_i)^2 }$
    \item Manhattan Distance - $d_{\text{manhattan}}(\mathbf{x}, \mathbf{y}) = \sum_{i=1}^n |x_i - y_i|$
    \item Chebyshev Distance - $d_{\text{chebyshev}}(\mathbf{x}, \mathbf{y}) = \max_{i} |x_i - y_i|$ 
    \item Cosine Distance - $d_{\text{cosine}}(\mathbf{x}, \mathbf{y}) = 1 - \frac{ \mathbf{x} \cdot \mathbf{y} }{ \|\mathbf{x}\| \cdot \|\mathbf{y}\| }$
    \item Bray-Curtis Dissimilarity - $d_{\text{braycurtis}}(\mathbf{x}, \mathbf{y}) = \frac{ \sum_{i=1}^n |x_i - y_i| }{ \sum_{i=1}^n (x_i + y_i) }$
\end{itemize}

\section*{Multi-Layer Degeneracy Index (MLDI)}

The goal is to quantify how much degeneracy exists across multiple layers of a communication stack. Let us define Layers given by, \( \mathcal{L} = \{\ell_1, \dots, \ell_k\} \). For each layer \( \ell_i \), let \( \mathcal{E}_{\ell_i} \) elements are performing a given function. Also, let \( \mathcal{D}_{\ell_i} \subset \mathcal{E}_{\ell_i} \) be the set of elements that are functionally equivalent and structurally diverse. Then,
\begin{align}
    \text{MLDI} = \frac{1}{k} \sum_{i=1}^{k} \frac{|\mathcal{D}_{\ell_i}|}{|\mathcal{E}_{\ell_i}|}
\end{align}

Basic MLDI is a ratio. Let us enhance it by a) modeling each layer’s degeneracy as entropy over capability distributions and b) including cross-layer redundancy and conditional entropy. Let \( \ell_i \in \mathcal{L} \) be layers of the protocol stack. Then, for each layer, \( \mathcal{E}_{\ell_i} = \{e^{(i)}_1, ..., e^{(i)}_{n_i}\} \). Also, for each entity, let us define the functionality vector \( \vec{f}^{(i)}_j \in \{0,1\}^m \) (binary indicator of what functions it can perform). Let us construct probability distribution over functions:
\begin{align}
    P^{(i)}(f_k) = \frac{1}{n_i} \sum_{j=1}^{n_i} \vec{f}^{(i)}_{jk}
\end{align}
Let the Layer Entropy be defined as,
\begin{align}
    H(\ell_i) = - \sum_{k=1}^m P^{(i)}(f_k) \log P^{(i)}(f_k)
\end{align}
where \( H(\ell_i \mid \ell_{i+1}) \) is the conditional entropy of function coverage across adjacent layers (e.g., PHY $\to$ MAC). In the above context, we can define an enhanced MLDI as,
\begin{align}
    \text{MLDI}^* = \frac{1}{k} \sum_{i=1}^k \left( \frac{H(\ell_i)}{\log m} + \gamma \cdot \frac{H(\ell_i \mid \ell_{i+1})}{\log m} \right)
\end{align}
where \( \gamma \in [0,1] \) is the weighting parameter and \( \log m \) normalizes entropy. Higher the $\text{MLDI}^*$, higher will be the system-wide functional diversity with cross-layer resilience.

\end{document}